\documentclass[11pt,a4paper]{article}
\usepackage{jheppub_kim}
\usepackage{rotating} 
\usepackage{graphicx}
\usepackage{amsmath}
\usepackage {amssymb}
\usepackage{subfigure}

\usepackage{relsize}
\providecommand{\U}[1]{\protect\rule{.1in}{.1in}}

\newcommand{\newc}{\newcommand}

\newc{\be}{\begin{equation}}
\newc{\ee}{\end{equation}}

\newc{\ba}{\begin{eqnarray}}
\newc{\ea}{\end{eqnarray}}
\newc{\bea}{\begin{eqnarray*}}
\newc{\eea}{\end{eqnarray*}}
\newc{\D}{\partial}
\newc{\ie}{{\it i.e.} }
\newc{\eg}{{\it e.g.} }
\newc{\etc}{{\it etc.} }
\newc{\etal}{{\it et al.}}
\newc{\lcdm}{$\Lambda$CDM }

\newc{\ra}{\Rightarrow}

\title{ Highly suppressed tensor-to-scalar ratio from a modified Lennard-Jones 
inflationary potential}
\author[a,b]{Panagiotis G. Stavros}

 \author[b,c,d]{Spyros~Basilakos}

\author[b,e,f]{Emmanuel N. Saridakis}
\affiliation[a]{Department of Physics, School of Sciences, University of 
Thessaly, 
3rd Km Old National Road, 35100, Lamia, Greece}

\affiliation[b]{Institute for Astronomy, Astrophysics, Space Applications and 
Remote Sensing, National Observatory of Athens, 15236 Penteli, Greece}
 
  \affiliation[c]{Academy of Athens, Research Center for Astronomy and Applied 
Mathematics, Soranou Efesiou 4, 11527, Athens, Greece}

 \affiliation[d]{School of Sciences, European University Cyprus, Diogenes 
Street, 
Engomi, 1516 Nicosia, Cyprus}

\affiliation[e]{Department of Astronomy, School of Physical Sciences, 
University of Science and Technology of China, Hefei 230026, P.R. China}

 \affiliation[f]{Departamento de Matem\'{a}ticas, Universidad Cat\'{o}lica del 
Norte, 
Avda.
Angamos 0610, Casilla 1280 Antofagasta, Chile}

\emailAdd{pstavros@uth.gr}
\emailAdd{svasil@academyofathens.gr} 
 \emailAdd{msaridak@noa.gr}

\abstract{  
 
The increasingly stringent observational bounds on primordial gravitational
waves strongly constrain inflationary model building, favoring scenarios that
predict highly suppressed tensor perturbations.
While many viable constructions rely on non-canonical kinetic terms,
non-minimal couplings, or modifications of gravity, it remains an open question
whether comparably small tensor amplitudes can emerge within a minimal,
single-field framework driven solely by potential dynamics.
In this work we propose a novel inflationary scenario based on a modified
Lennard-Jones potential.
Inspired by a well-known interaction potential in molecular physics, the
proposed form naturally combines a smooth minimum with an extended flat plateau
at large field values.
This intrinsic structure supports slow-roll inflation  and ensures a graceful 
exit without introducing additional degrees of freedom.
We perform a detailed analysis of the inflationary dynamics and confront the
model with current observational constraints.
We find that the scalar spectral index is fully consistent with CMB data, while
the tensor-to-scalar ratio is predicted to be extremely small, reaching values
as low as $r\sim10^{-7}$. Finally, the running of the scalar spectral index is 
also found to be small, well withing the 1$\sigma$ recent observational bounds 
from Atacama Cosmology 
Telescope.

}
\begin{document}
\maketitle
\section{Introduction}

Inflation constitutes one of the most successful paradigms in modern cosmology
\cite{Kazanas:1980tx,Sato:1980yn,Guth:1980zm,Linde:1981mu},
providing a compelling explanation for the observed homogeneity, isotropy and
near flatness of the Universe, while simultaneously offering a natural mechanism
for the generation of primordial density perturbations.
Its generic predictions, and in particular the emergence of an almost
scale-invariant spectrum of scalar fluctuations, are in remarkable agreement
with high-precision observations of the cosmic microwave background (CMB).
Nevertheless, despite this phenomenological success, the fundamental origin of
inflation and the detailed form of the dynamics driving the accelerated phase
remain open problems, motivating the continuous development and refinement of
inflationary models.

>From a theoretical point of view, inflationary scenarios can be broadly
classified into models driven by scalar-field dynamics and models emerging from
modifications of the gravitational sector 
\cite{Olive:1989nu,Lyth:1998xn,Bartolo:2004if,Nojiri:2010wj,
Martin:2013tda},
leading to a wide spectrum of viable constructions capable of producing a
sufficiently long phase of accelerated expansion.
However, the steadily improving observational bounds now act as precision
filters, excluding large classes of models and severely constraining the
inflationary parameter space.

The most relevant observables constraining inflation include the amplitude of
the scalar power spectrum $A_s$, the scalar spectral index $n_s$, the
tensor-to-scalar ratio $r$, and possibly the running of the scalar spectral
index $\alpha_s \equiv d n_s/d\ln k$.
Current data fix $A_s \sim 10^{-9}$ and indicate a slightly red-tilted scalar
spectrum with $n_s \simeq 0.965$, implying that inflation must occur in a regime
where the slow-roll conditions are well satisfied.
Among these quantities, the tensor-to-scalar ratio plays a particularly central
role, since it directly probes the inflationary energy scale and the production
of primordial gravitational waves.

Recent results from Planck and the BICEP/Keck Array
\cite{Planck:2018jri,BICEP:2021xfz} impose a stringent upper bound
$r \lesssim 0.036$ at $95\%$ confidence level.
This constraint has far-reaching consequences for inflationary model building,
as it strongly disfavors simple large-field monomial potentials, which typically
predict sizeable tensor amplitudes.
As a consequence, significant attention has shifted toward inflationary models
characterized by plateau-like potentials at large field values, where the
slow-roll parameter $\varepsilon_V$ is strongly suppressed and the tensor
contribution becomes naturally small.
Such behavior arises in a variety of frameworks, including non-minimally
coupled scalar-field models 
\cite{Faraoni:1996rf,Geng:2017mic,Benisty:2019jqz,Panda:2022can,
Basilakos:2023xof,Bastero-Gil:2006zpr,Einhorn:2009bh,Bauer:2008zj,
Germani:2010gm,Feng:2010ya,Hertzberg:2010dc,Hossain:2014xha,Hossain:2014coa,
Hossain:2014ova,Geng:2015fla,Lola:2020lvk,Braglia:2020fms,Karydas:2021wmx,
Basilakos:2015sza,
Papanikolaou:2022did,Sohail:2024oki,Nojiri:2015fra}, Horndeski and Galileon 
theories \cite{Tsujikawa:2014mba,Germani:2015plv,BeltranJimenez:2017cbn,
Sebastiani:2017cey,Oikonomou:2020sij,Chen:2021nio,
Kobayashi:2010cm,Burrage:2010cu,Renaux-Petel:2011lur,Renaux-Petel:2011rmu,
Choudhury:2012whm,Ohashi:2012wf,Qiu:2015nha,Choudhury:2023hvf,
Choudhury:2023hfm}, as well as 
modified
gravity constructions 
\cite{Garcia-Bellido:1995him,Paliathanasis:2017apr,Nojiri:2017ncd,Awad:2017ign,
Chakraborty:2018scm, 
Qiu:2018nle,Shimada:2018lnm,Gamonal:2020itt,Cai:2021uup,Papanikolaou:2021uhe, 
Shiravand:2022ccb,Tzerefos:2023mpe,Nojiri:2007bt,Bamba:2008ja,CANTATA:2021ktz, 
Barrow:1988xh,Nojiri:2003ft,Carter:2005fu,Ferraro:2006jd,Nojiri:2007as, 
Nojiri:2007cq,Mukohyama:2009gg,DeLaurentis:2015fea,Bamba:2016wjm,Nozari:2010kri,
 Elizalde:2010ep,Cai:2010kp,Briscese:2012ys,Sebastiani:2013eqa,Bamba:2014jia, 
Bamba:2014mua,Nojiri:2014zqa,Bhat:2023qwa,Sadatian:2024lub,
Capozziello:2022tvv}.     A particularly well-known example is Starobinsky $R^2$ 
inflation
\cite{Starobinsky:1980te}, which predicts a naturally suppressed
tensor-to-scalar ratio.

Motivated by the fact that current observations remain fully compatible with
$r=0$, a growing class of inflationary scenarios has been developed with the
explicit aim of producing parametrically small tensor amplitudes.
Among them, attractor-type models constitute a prominent category, where
non-canonical kinetic structures lead, after canonical normalization, to
universal plateau-like potentials
\cite{Kallosh:2013hoa,Kallosh:2013yoa,Kallosh:2013tua,Galante:2014ifa,
Linde:2015uga,Gao:2017uja,Miranda:2017juz,Dimopoulos:2017zvq,Dimopoulos:2017tud,
Scalisi:2018eaz,Braglia:2020bym,Bhattacharya:2022akq,Brissenden:2023yko,
Ferrara:2016fwe,Akarsu:2016qhf}.  
Although phenomenologically successful, such constructions typically rely on
specific kinetic terms or non-minimal ingredients, raising the question of
whether comparably suppressed tensor-to-scalar ratios can be achieved within
simpler, minimally coupled single-field frameworks.

In this work we demonstrate that this is indeed possible, and we propose a 
novel inflationary scenario based on a modified Lennard-Jones
potential.
Inspired by a familiar interaction potential in molecular physics, the proposed
form naturally combines a smooth minimum with an extended, flat plateau at
large field values.
Remarkably, this simple structure is sufficient to support slow-roll inflation
within a minimally coupled, single-field setup, without invoking non-canonical
kinetic terms or modifications of gravity.
We show that the resulting inflationary dynamics yield scalar spectral indices
fully consistent with current observations, while leading to an exceptionally
suppressed tensor-to-scalar ratio, which can reach values as small as
$r \sim 10^{-7}$.
This suppression arises directly from the intrinsic shape of the potential and
its plateau structure, rather than from additional theoretical ingredients.

The paper is organized as follows.
In Section~\ref{model} we introduce the modified Lennard-Jones potential and
discuss its basic properties.
In Section~\ref{inflationary_dynamics} we analyze the inflationary evolution,
derive the slow-roll parameters and inflationary observables, and confront the
model with current observational constraints, while we also discuss the 
reheating realization.  
Finally, Section~\ref{Conclusions} summarizes the main results and outlines
possible directions for future investigation.

\section{The modified Lennard-Jones inflationary model}
\label{model} 

In this section we introduce the modified Lennard-Jones inflationary model and
outline its basic theoretical setup.
We first briefly recall the standard inflationary framework in which our model
is embedded, and then present the explicit form of the modified Lennard-Jones
potential that drives inflation.
The resulting dynamics and their observational implications are analyzed in
detail in the following subsections.

\subsection{Basic inflationary framework}

We consider a single real scalar field $\phi$ minimally coupled to gravity, 
described by the action
\begin{equation}
S=\int d^{4}x\,\sqrt{-g}\left[\frac{M_{\rm Pl}^{2}}{2}R
-\frac{1}{2}g^{\mu\nu}\partial_{\mu}\phi\,\partial_{\nu}\phi
- V(\phi)\right],
\label{action}
\end{equation}
where $M_{\rm Pl}\simeq 2.4\times10^{18}\,{\rm GeV}$ is the reduced Planck mass.
Assuming a spatially flat Friedmann-Robertson-Walker metric
\begin{equation}
ds^{2}=-dt^{2}+a^{2}(t)\delta_{ij}dx^{i}dx^{j},
\end{equation}
and a homogeneous scalar field $\phi=\phi(t)$, variation of the action yields 
the Friedmann equations
\begin{equation}
3M_{\rm Pl}^{2}H^{2}=\frac{1}{2}\dot{\phi}^{2}+V(\phi),
\qquad
\dot{H}=-\frac{1}{2M_{\rm Pl}^{2}}\dot{\phi}^{2},
\label{friedmann}
\end{equation}
together with the Klein-Gordon equation
\begin{equation}
\ddot{\phi}+3H\dot{\phi}+V'(\phi)=0,
\label{kg}
\end{equation}
where $H=\dot{a}/a$ is the Hubble function and a prime denotes differentiation 
with respect to $\phi$.

Inflation occurs when the potential energy dominates over the kinetic term, 
allowing for the slow-roll approximation. In this regime the dynamics is 
conveniently characterized by the potential slow-roll 
parameters \cite{Martin:2013tda}
\begin{eqnarray}
&&\varepsilon_V \equiv \frac{M_{Pl}^2}{2}\left(\frac{V'}{V}\right)^2,
\nonumber\\
&&
\eta_V \equiv M_{Pl}^2\frac{V''}{V},
\nonumber\\
&&
\xi_V^2 \equiv M_{Pl}^4\left(\frac{V'V'''}{V^2}\right),
\label{slowrollparametersLJ}
\end{eqnarray}
where primes denote differentiation with respect to $\phi$, 
with successful inflation requiring $\varepsilon_{V},|\eta_{V}|\ll1$.
To leading order in slow roll, the Hubble parameter and scalar-field evolution 
satisfy
\begin{equation}
H^{2}\simeq\frac{V}{3M_{\rm Pl}^{2},
}
\qquad
3H\dot{\phi}\simeq -V'(\phi).
\end{equation}
Additionally, quantum fluctuations of the inflaton generate primordial 
curvature 
perturbations 
with power spectrum
\begin{equation}
\Delta_{\mathcal{R}}^{2}
\simeq
\frac{1}{12\pi^{2}}
\frac{V^{3}}{M_{\rm Pl}^{6}\,V'^{2}}
\Big|_{k=aH},
\label{pscalar}
\end{equation}
evaluated at horizon crossing. 

The inflationary observables, such as the tensor-to-scalar ratio 
($r$), the scalar spectral
index ($n_s$) and its
running ($\alpha_s={d}n_s/{ d}\ln k$), as well as the tensor spectral index 
$n_T$ can be expressed in terms of the 
slow-roll 
parameters as \cite{Martin:2013tda}
\begin{eqnarray}
n_s &\approx& 1-6\varepsilon_V+2\eta_V,
\label{nsLJ}\\
\alpha_s &\approx& 16\varepsilon_V\eta_V-24\varepsilon_V^2-2\xi_V^2,
\label{alphasLJ}\\
r &\approx& 16\varepsilon_V,
\label{rLJ}\\
n_T &\approx& -2\varepsilon_V.
\label{ntLJ}
\end{eqnarray}
Additionally, using the slow-roll parameters, (\ref{pscalar}) can provide the 
potential at 
horizon crossing, namely
\begin{equation}
    V_{*}\equiv V( \phi_{*})=\frac{3\pi^2}{2}rA_{s}M_{\rm Pl}^4,
\label{vstar}
\end{equation}
with   $\phi_{*}$ the corresponding $\phi$ value.
In the following, these general expressions will be applied to a specific 
inflationary potential, whose form determines the detailed phenomenology.

\subsection{The modified Lennard-Jones potential}

We now introduce the inflationary potential that constitutes the core novelty 
of 
the present work. 
Inspired by the well-known Lennard-Jones potential describing short-range 
repulsion and long-range attraction in intermolecular interactions, we 
construct 
a modified form suitable for cosmological inflation. 
Our aim is to obtain a potential that (i) possesses a smooth minimum at the 
origin, allowing for a graceful exit from inflation, and (ii) exhibits an 
extended flat region at large field values, capable of supporting a prolonged 
slow-roll phase.

We start from the general shifted Lennard-Jones form
\begin{equation}
V(x)=M\left[\left(\frac{\beta}{x+x_{0}}\right)^{12}
-\left(\frac{\beta}{x+x_{0}}\right)^{6}\right]+c,
\end{equation}
where $\beta$ sets the characteristic scale of the interaction and $M$ 
determines 
the overall energy scale.
In order for the origin to correspond to a physical minimum of the potential, 
we 
impose the conditions
\begin{equation}
V(0)=0,
\qquad
V'(0)=0,
\end{equation}
and these requirements uniquely fix the shift and additive constant as
\begin{equation}
x_{0}=2^{1/6}\beta,
\qquad
c=\frac{M}{4}.
\end{equation}

Promoting the variable $x$ to the inflaton field $\phi$, we finally arrive at 
the modified Lennard-Jones inflationary potential
\begin{equation}
V(\phi)=V_{0}\left[
\left(\frac{\beta}{\phi+2^{1/6}\beta}\right)^{12}
-\left(\frac{\beta}{\phi+2^{1/6}\beta}\right)^{6}
+\frac{1}{4}
\right],
\label{LJpotential}
\end{equation}
where $V_{0}$ sets the height of the potential and $\beta$ controls the width 
of the 
transition between the minimum and the plateau. Hence, $\beta$ has dimensions 
of energy while $V_{0}$ as usual has dimensions 
of energy to the fourth.

The potential (\ref{LJpotential}) exhibits a quadratic minimum around $\phi=0$, 
ensuring coherent oscillations after inflation, while for large field values it 
asymptotically approaches a constant plateau. 
This combination naturally leads to small slow-roll parameters during the 
inflationary phase and hence to suppressed tensor-to-scalar ratios, in close 
analogy with plateau-type inflationary models.
However, an important distinction is that in the present scenario the 
flattening 
mechanism arises purely from the intrinsic structure of the modified 
Lennard-Jones form, without invoking non-minimal couplings or non-canonical 
kinetic terms.

In summary, the potential (\ref{LJpotential}) smoothly interpolates between a 
well-defined minimum and a flat large-field regime, providing a minimal and 
self-contained realization of single-field inflation.
In the following subsection we investigate in detail the corresponding 
inflationary dynamics and observational predictions.

\section{Inflationary dynamics under Lennard-Jones potential}
\label{inflationary_dynamics}

 In this section we perform a detailed quantitative analysis of the inflationary
dynamics driven by the modified Lennard-Jones potential.
We examine the slow-roll evolution, determine the duration of inflation, and
calculate the corresponding inflationary observables at horizon crossing.
The predictions of the model are then confronted with current observational
constraints.

\subsection{Inflationary realization}

Let us now perform a quantitative analysis of the inflationary evolution driven 
by the modified Lennard-Jones potential (\ref{LJpotential}). Since we are 
within a standard minimally-coupled single-field framework, we can use the 
slow-roll definitions (\ref{slowrollparametersLJ}).

For the potential (\ref{LJpotential}) we obtain 
\begin{equation}
\frac{V'(\phi)}{V(\phi)}=
\frac{
6 \beta^6 \left[\left(2^{1/6} \beta+\phi \right)^6-2 \beta^6\right]
}{
\left(2^{1/6} \beta+\phi \right)^{13}
\left[
\frac{\beta^6\left(\beta^6-\left(2^{1/6} 
\beta+\phi\right)^6\right)}{\left(2^{1/6} 
\beta+\phi\right)^{12}}
+\frac{1}{4}
\right]
}.
\label{VoverV}
\end{equation}
Inflation ends when the slow-roll conditions break down, i.e. when one of the 
slow-roll parameters becomes of order unity. In practice, and as usual, we 
determine the end of inflation by the condition $\varepsilon_V(\phi_{\rm 
end})=1$, namely
\begin{equation}
\frac{M_{Pl}^{2}}{2}\left(\frac{V'}{V}\right)^{2}\Bigg|_{\phi=\phi_{\rm end}}=1,
\label{endinfl}
\end{equation}
which yields $\phi_{\rm end}$ numerically for each parameter choice.
Finally, the total number of e-foldings between the onset of inflation at 
$\phi_*$ and 
the end of inflation at $\phi_{\rm end}$ is given by
\begin{equation}
N=\int_{t_*}^{t_{\rm end}} H dt
\simeq
\frac{1}{M_{Pl}^{2}}\int_{\phi_{\rm end}}^{\phi_*}\frac{V}{V'}\,d\phi,
\label{efolds}
\end{equation}
where in the last equality we used the slow-roll approximation. For the 
phenomenological analysis we will focus on the representative values $N=50$,  
$N=55$ and $N=60$.

>From the inflationary observables expressions (\ref{nsLJ})-(\ref{ntLJ}), and 
using the above Lennard-Jones potential,  we see that   the model at hand can 
successfully accommodate the current observational
constraints on the scalar spectral index and the tensor-to-scalar ratio.
In particular, requiring compatibility with $n_s \simeq 0.97$ and the
Planck/BICEP bound $r<0.044$, we obtain viable inflationary trajectories for
parameter values of order $
\beta \simeq 10^{16}- 10^{19}\,\text{GeV} $.
In all cases inflation starts at super-Planckian field values,
$\phi_*>\sqrt{2}M_{\rm Pl}$, and terminates at sub-Planckian values.

\begin{figure}[ht]
\centering
\includegraphics[width=0.70\linewidth]{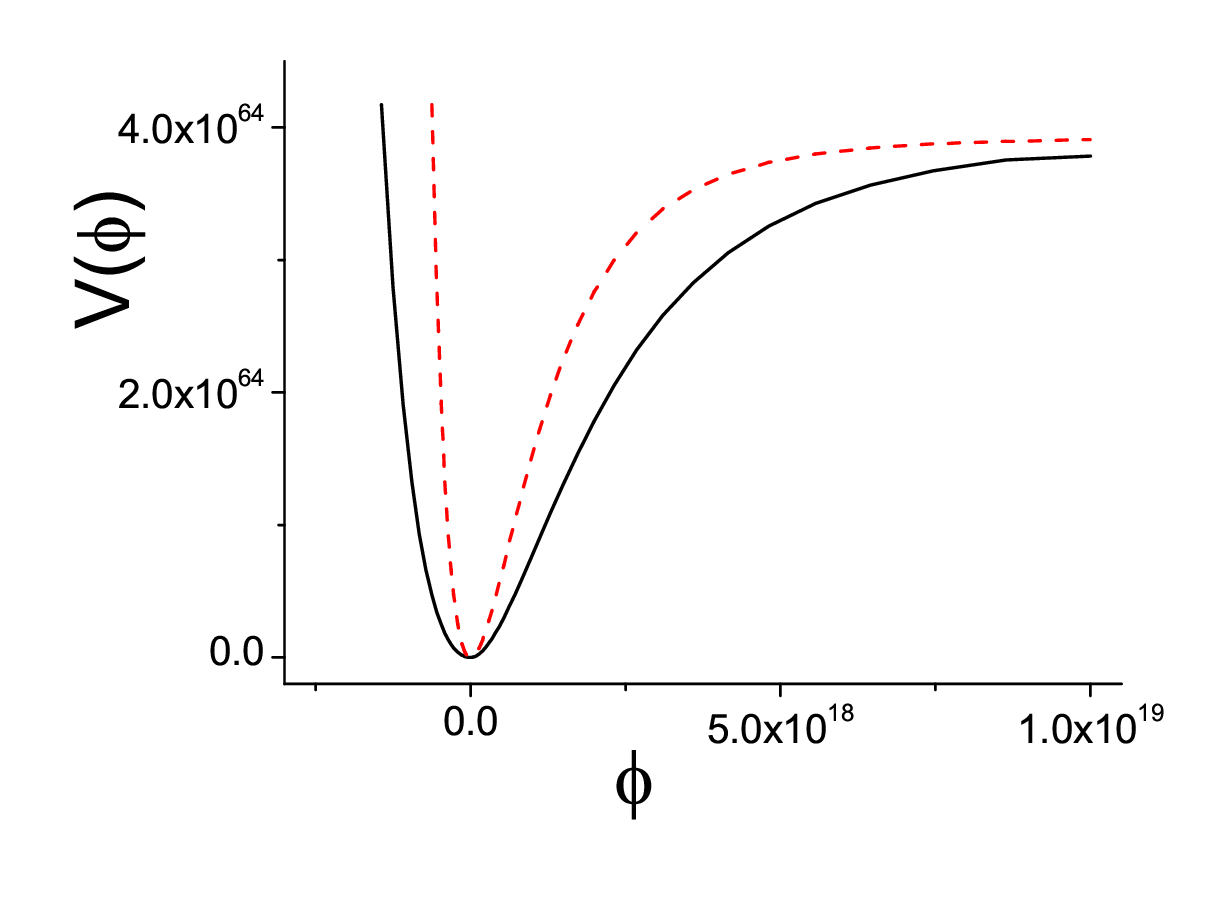}
\caption{{\it{The modified Lennard-Jones potential $V(\phi)$ of 
Eq.~(\ref{LJpotential}) for $V_0 \simeq  10^{65}\,{\rm GeV}^4$, and for 
$\beta= 10^{19}\,{\rm GeV}$ (black-solid curve) and $\beta=5\times 
10^{18}\,{\rm GeV}$ (red-dashed curve).}}}
\label{fig:V}
\end{figure}

Fixing the end of inflation through the condition $\varepsilon_V(\phi_{\rm 
end})=1$
and imposing the observed amplitude of scalar perturbations,
$A_s\simeq 2.19\times 10^{-9}$ at horizon crossing, using (\ref{vstar}) we find 
for the 
representative
case $N=60$ that
\begin{equation}
V_0 \simeq 1.57\times 10^{65}\,{\rm GeV}^4 ,
\end{equation}
with $\phi_{*}=2.3\times 10^{19}\,{\rm GeV}$. Moreover, for the 
corresponding Hubble function  we obtain $H_{*}\approx 4.24 \times 
10^{13}\,{\rm 
GeV} $ which is in agreement with the standard lore.
Finally, in order to be more transparent, in Fig.~\ref{fig:V} we depict the 
modified Lennard-Jones potential for   
$V_0= 10^{65}$GeV$^4$, and for   two values of the potential parameter 
$\beta$.
As can be seen, the potential exhibits a smooth minimum at $\phi=0$ and develops
a plateau at large field values, which is the key ingredient responsible for the
slow-roll regime.   

  \begin{figure}[ht]
\centering
\hspace{-1.cm}
\includegraphics[scale=.48]{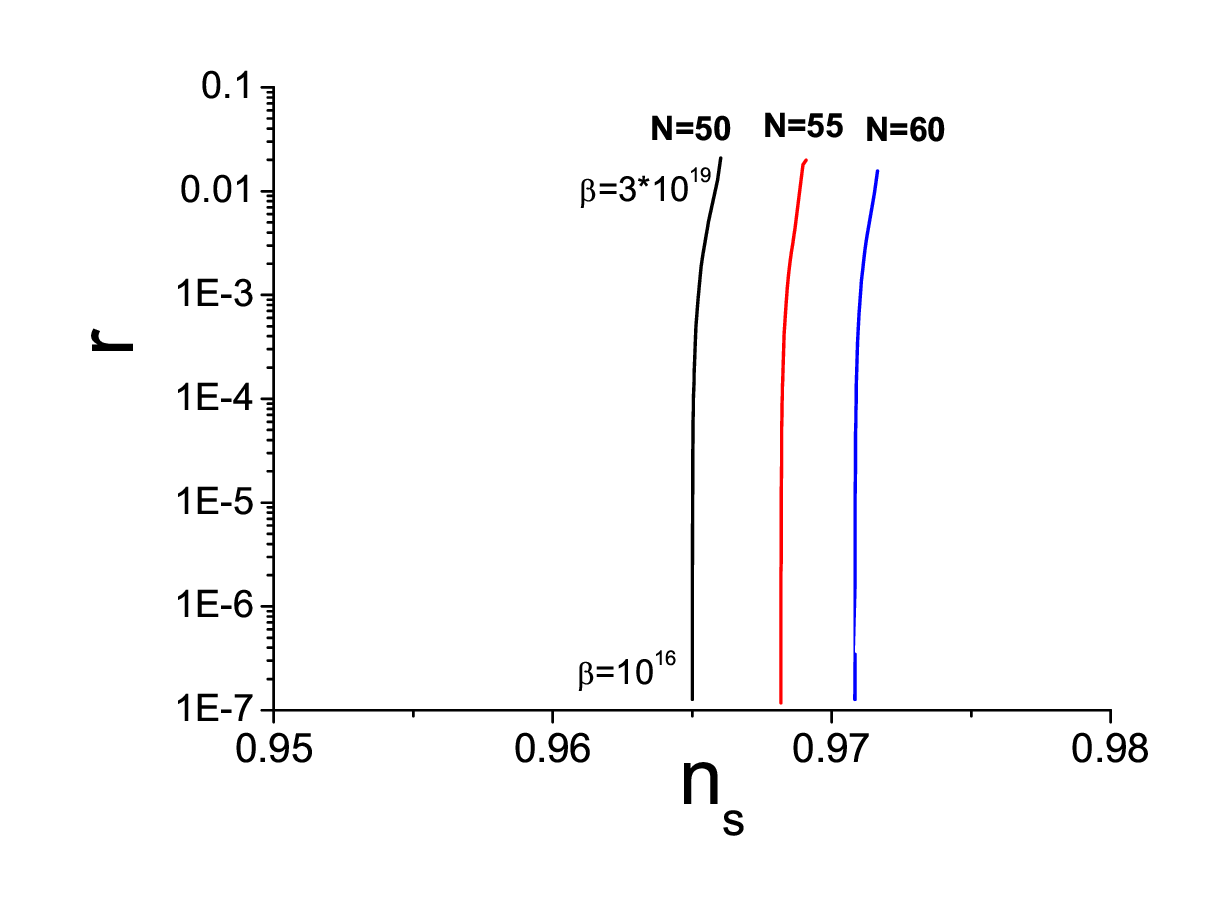}
\caption{{\it{ The predictions of the inflationary scenario based on the 
 modified Lennard-Jones potential (\ref{LJpotential})  on the  
$n_{\mathrm{s}}-r$ plane. The left     curve corresponds to 
e-folding number $N=50$, with   $\beta$ varying from $10^{16}$ GeV to 
$3\times 10^{19}$ GeV,  while the middle and right curves correspond
to $N=55$ and $N=60$, respectively, with the same variance of   $\beta$. In 
all curves  $V_0$ is set around 
$V_0= 10^{65}$GeV$^4$ in order to obtain   $A_s = 2 \times 10^{-9}$. 
}}}
\label{rns}
\end{figure}

\begin{figure}[ht]
\centering
\hspace{-1.cm}
\includegraphics[scale=.55]{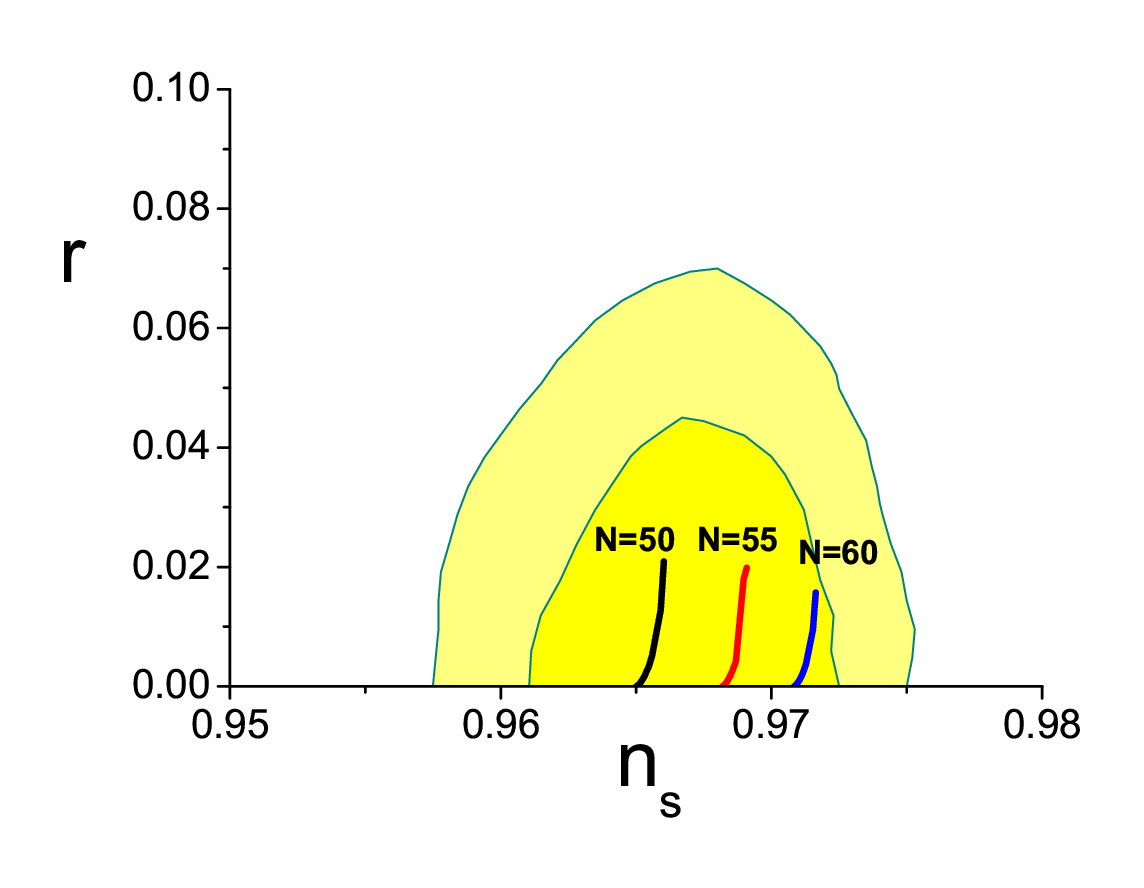}
\caption{{\it{The predictions of Fig. \ref{rns} on top of the 1$\sigma$  and 
2$\sigma$   Planck 2018  TT,TE,EE+lowE+lensing +BK15+BAO results 
\cite{Planck:2018jri}.
}}}
\label{rnsdata}
\end{figure}

 Let us proceed to the investigation of the inflationary dynamics, evaluated   
at the horizon crossing. 
In Fig. \ref{rns} we depict the obtained 
tensor-to-scalar ratio $r$ versus the spectral index  $n_s$, with the model 
parameter $\beta$ varying from $10^{16}$ GeV to 
$3\times 10^{19}$ GeV.
  As we can see, as  $\beta$ decreases,   
  $r$ acquires     smaller  values, that can reach up to
$10^{-7}$. These values are   significantly smaller than in quadratic chaotic 
inflation, which  reveals the efficiency of the  Lennard-Jones potential.
In order to be more transparent,  
    in  Fig. \ref{rnsdata}  we display the predictions of Fig. 
\ref{rns} on top of the  Planck 2018  observational  results, and as we observe 
the agreement is excellent, since the model predictions lie well within the 
1$\sigma$ region.

\begin{figure}[ht]
\centering
\includegraphics[scale=.55]{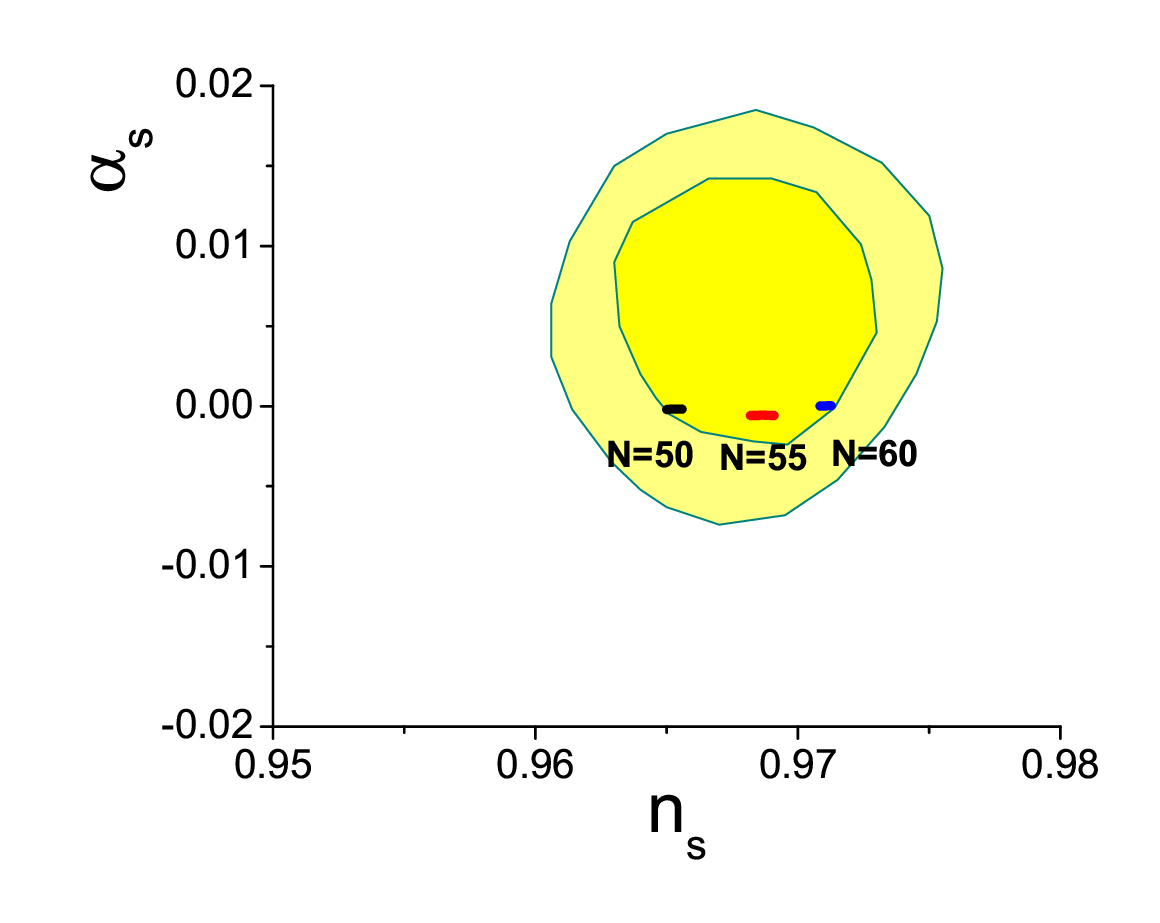}
\caption{{\it{The predictions of the inflationary scenario based on the 
 modified Lennard-Jones potential (\ref{LJpotential})  on the  
running spectral index $\alpha_s$. The left     curve corresponds to 
e-folding number $N=50$, with   $\beta$ varying from $10^{16}$ GeV to 
$3\times 10^{19}$ GeV, while the middle and right curves corresponds
to $N=55$ and $N=60$, respectively, with the same variance of 
    $\beta$. In all curves  $V_0$ is set around 
$V_0= 10^{65}$GeV$^4$ in order to obtain   $A_s = 2 \times 10^{-9}$. 
Additionally, we present the 1$\sigma$ (yellow) and 2$\sigma$ (light yellow) 
contours for the Atacama Cosmology Telescope (P-ACT)
results   \cite{ACT:2025tim}.}}}
\label{runningspectral}
\end{figure}

Finally, let us examine the running of the specral index $\alpha_s$. Using  
(\ref{alphasLJ})    we can calculate it for the above values of 
  $\beta$  and $N$, and  in Fig. \ref{runningspectral} we 
present the results on the 
$\alpha_s-n_s$ plane. On the same graph we add 
the 1$\sigma$   and 2$\sigma$ 
contours for the Atacama Cosmology Telescope (P-ACT)
results   \cite{ACT:2025tim}. As we can see, the predictions of the model lie 
within the 1$\sigma$ observational contour, which is an additional advantage.
 
 In summary, from the analysis it becomes clear that the inflationary scenario 
based 
on the  modified Lennard-Jones potential reproduces
successfully  the observed   scalar power spectrum, both its amplitude and 
running, while it can predict very small values of the tensor-to-scalar ratio.

\subsection{Reheating considerations}

After the end of inflation, the scalar field settles into coherent oscillations
around the minimum of the potential at $\phi=0$.
As discussed above, the modified Lennard-Jones potential possesses a smooth
quadratic minimum, which naturally allows for a standard reheating mechanism
through perturbative inflaton decay within a minimal setup.

Expanding the potential around its minimum, the inflaton mass is determined by
the second derivative of the potential evaluated at $\phi=0$, namely
\begin{equation}
m_{\phi} = \sqrt{V''(0)} \simeq 1.9\times 10^{13}\,{\rm GeV}.
\end{equation}
Assuming that reheating proceeds through gravitationally suppressed decay
channels, the corresponding decay rate is estimated as
\begin{equation}
\Gamma \sim \frac{m_{\phi}^{3}}{M_{\rm Pl}^{2}}
\simeq 1.6\times 10^{3}\,{\rm GeV}.
\end{equation}
The reheating temperature is then given by
\begin{equation}
T_{\rm reh} \sim
\left(\frac{90}{\pi^{2}g_{*}}\right)^{1/4}
\sqrt{\Gamma M_{\rm Pl}}
\simeq 3.14\times 10^{10}\,{\rm GeV},
\end{equation}
where $g_{*}\sim 100$ denotes the effective number of relativistic degrees of
freedom.

This value comfortably satisfies the Big Bang Nucleosynthesis bound,
$T_{\rm reh}\gtrsim 1\,{\rm MeV}$, ensuring that the synthesis of light elements
remains unaffected.
Moreover, the resulting reheating temperature lies well within the range
commonly encountered in minimal single-field inflationary scenarios with
gravitationally suppressed couplings, indicating that no additional reheating
ingredients are required for phenomenological viability.
Possible direct couplings of the inflaton to matter fields could further modify
the reheating dynamics, but such extensions are not necessary for the internal
consistency of the present framework.

\section{Conclusions}
\label{Conclusions}

In this work we proposed and investigated a novel inflationary scenario based on
a modified Lennard-Jones potential.
The motivation behind this construction was twofold.
On the one hand, current observations impose increasingly stringent constraints
on the tensor-to-scalar ratio, strongly disfavoring simple large-field
inflationary models and pushing viable scenarios toward plateau-like potentials.
On the other hand, many mechanisms that successfully suppress tensor modes rely
on non-canonical kinetic terms, non-minimal couplings, or modifications of the
gravitational sector, which naturally raises the question of whether 
similarly successful predictions
can be achieved within a minimal, single-field framework, with standard kinetic
terms and Einstein gravity.
The present work provides a positive answer to this question.

The proposed modified Lennard-Jones potential exhibits a smooth minimum at the
origin and a flat plateau at large field values.
This intrinsic structure is sufficient to support slow-roll inflation and to
ensure a graceful exit through coherent oscillations around the minimum.
We performed a detailed analysis of the inflationary dynamics, deriving the
slow-roll parameters, the number of e-folds, and the corresponding inflationary
observables.
We found that the model naturally yields scalar spectral indices fully
consistent with current observational bounds, while predicting a remarkably
suppressed tensor-to-scalar ratio.
In particular, the tensor amplitude can reach values as small as
$r \sim 10^{-7}$, significantly below those of simple chaotic models and even
below the typical predictions of standard plateau scenarios.
This strong suppression arises directly from the shape of the potential and the
rapid decay of the slow-roll parameter $\varepsilon_V$ on the plateau, rather
than from fine-tuning or additional theoretical ingredients.
Furthermore, the model predicts a small running of the scalar spectral index,
with $|\alpha_s| \ll 1$, lying well within the current observational 
constraints.

An important feature of the present scenario is its conceptual simplicity.
The inflaton field is minimally coupled to gravity and governed by a standard
canonical kinetic term, implying that the inflationary predictions are entirely
driven by the shape of the potential.
In this sense, the model provides a simple and self-contained example in which
extremely small tensor modes emerge naturally from the potential dynamics alone,
without invoking non-canonical kinetic terms, non-minimal couplings, or
additional degrees of freedom.
Furthermore, the presence of a smooth quadratic minimum ensures a graceful exit
from inflation through coherent oscillations of the inflaton field.

We also examined the reheating phase within this minimal framework and found
that it proceeds efficiently.
The resulting reheating temperature is comfortably within the Big Bang
Nucleosynthesis bound, indicating that the post-inflationary thermal history of
the Universe remains fully viable without the need for additional interactions
or exotic decay channels.
This result further reinforces the internal consistency of the model, showing
that both the inflationary and reheating phases can be successfully realized
within a single-field, minimally coupled setup.

The results presented here open several directions for future investigation.
A natural extension of the present work would be to explore generalized
Lennard-Jones-type potentials and assess the robustness of the strong tensor
suppression against deformations of the plateau structure. 
Additionally, given the extremely small tensor-to-scalar ratio predicted by this
scenario, the model provides a well-motivated benchmark for future CMB
experiments, such as LiteBIRD and CMB-S4, whose continued null detections of
primordial tensor modes could further discriminate between minimal plateau
models and more elaborate inflationary constructions. These studies are left 
for future projects.

\begin{acknowledgments}  
We would like to thank George Pavlopoulos for
insightful discussions. The authors acknowledge the contribution of the LISA CosWG, and of   COST 
Actions   CA21136 ``Addressing observational tensions in cosmology with 
systematics and fundamental physics (CosmoVerse)'', CA21106 ``COSMIC WISPers 
in the Dark Universe: Theory, astrophysics and experiments'', 
and CA23130 ``Bridging high and low energies in search of quantum gravity 
(BridgeQG)''.

\end{acknowledgments}

%%%%%%%%%%%%%%%%%%%%%%%%%%%%% References 
%%%%%%%%%%%%%%%%%%%%%%%%%%%%%%%%%%%%%%%%%


\begin{thebibliography}{}



\bibitem{Kazanas:1980tx}
  D.~Kazanas,
  {\it{Dynamics of the Universe and Spontaneous Symmetry Breaking}},
  Astrophys. J. Lett. \textbf{241}, L59-L63 (1980).

\bibitem{Sato:1980yn}
  K.~Sato,
  {\it{First Order Phase Transition of a Vacuum and Expansion of the Universe}},
  Mon.\ Not.\ Roy.\ Astron.\ Soc.\  {\bf 195}, 467 (1981).

\bibitem{Guth:1980zm}
  A.~H.~Guth,
  {\it{The Inflationary Universe: A Possible Solution to the Horizon and 
Flatness Problems}},
  Phys.\ Rev.\ D {\bf 23}, 347 (1981).

\bibitem{Linde:1981mu}
  A.~D.~Linde,
  {\it{A New Inflationary Universe Scenario: A Possible Solution of the 
Horizon, 
Flatness, Homogeneity, Isotropy and Primordial Monopole Problems}},
  Phys.\ Lett.\ B {\bf 108}, 389 (1982).

\bibitem{Olive:1989nu}
  K.~A.~Olive,
  {\it{Inflation}},
  Phys.\ Rept.\  {\bf 190}, 307 (1990).

\bibitem{Lyth:1998xn}
  D.~H.~Lyth and A.~Riotto, 
  {\it{Particle physics models of inflation and the cosmological density 
perturbation}}, 
  Phys.\ Rept.\  {\bf 314}, 1 (1999) 
  [\href{https://arxiv.org/abs/hep-ph/9807278}{{\tt arXiv:hep-ph/9807278}}].

 

 

 




 
 
 \bibitem{Bartolo:2004if}
  N.~Bartolo, E.~Komatsu, S.~Matarrese and A.~Riotto,
  {\it{Non-Gaussianity from inflation: Theory and observations}},
  Phys.\ Rept.\  {\bf 402}, 103 (2004)
  [\href{https://arxiv.org/abs/astro-ph/0406398}{{\tt arXiv:astro-ph/0406398}}].

\bibitem{Nojiri:2010wj}
  S.~Nojiri and S.~D.~Odintsov,
  {\it{Unified cosmic history in modified gravity: from F(R) theory to Lorentz 
non-invariant models}},
  Phys.\ Rept.\  {\bf 505}, 59-144 (2011)
  [\href{https://arxiv.org/abs/1011.0544}{{\tt arXiv:1011.0544}}].

\bibitem{Martin:2013tda}
  J.~Martin, C.~Ringeval and V.~Vennin,
  {\it{Encyclopaedia Inflationaris}},
  Phys.\ Dark\ Univ.\  {\bf 5-6}, 75-235 (2014)
  [\href{https://arxiv.org/abs/1303.3787}{{\tt arXiv:1303.3787}}].

\bibitem{BICEP:2021xfz}
  P.~A.~R.~Ade {\it{et al.}} [BICEP and Keck],
  {\it{Improved Constraints on Primordial Gravitational Waves using Planck, 
WMAP, and BICEP/Keck Observations through the 2018 Observing Season}},
  Phys.\ Rev.\ Lett.\  {\bf 127}, 151301 (2021)
  [\href{https://arxiv.org/abs/2110.00483}{{\tt arXiv:2110.00483}}].

\bibitem{Planck:2018jri}
  Y.~Akrami {\it{et al.}} [Planck],
  {\it{Planck 2018 results. X. Constraints on inflation}},
  Astron.\ Astrophys.\  {\bf 641}, A10 (2020)
  [\href{https://arxiv.org/abs/1807.06211}{{\tt arXiv:1807.06211}}].

    
\bibitem{Faraoni:1996rf}
  V.~Faraoni,
  {\it{Nonminimal coupling of the scalar field and inflation}},
  Phys.\ Rev.\ D  {\bf 53}, 6813-6821 (1996)
  [\href{https://arxiv.org/abs/astro-ph/9602111}{{\tt arXiv:astro-ph/9602111}}].

\bibitem{Bastero-Gil:2006zpr}
  M.~Bastero-Gil, S.~F.~King and Q.~Shafi,
  {\it{Supersymmetric Hybrid Inflation with Non-Minimal Kahler potential}},
  Phys.\ Lett.\ B  {\bf 651}, 345-351 (2007)
  [\href{https://arxiv.org/abs/hep-ph/0604198}{{\tt arXiv:hep-ph/0604198}}].


\bibitem{Bauer:2008zj}
  F.~Bauer and D.~A.~Demir,
  {\it{Inflation with Non-Minimal Coupling: Metric versus Palatini 
Formulations}},
  Phys.\ Lett.\ B  {\bf 665}, 222-226 (2008)
  [\href{https://arxiv.org/abs/0803.2664}{{\tt arXiv:0803.2664}}].



\bibitem{Einhorn:2009bh}
  M.~B.~Einhorn and D.~R.~T.~Jones,
  {\it{Inflation with Non-minimal Gravitational Couplings in Supergravity}},
  JHEP  {\bf 03}, 026 (2010)
  [\href{https://arxiv.org/abs/0912.2718}{{\tt arXiv:0912.2718}}].

\bibitem{Germani:2010gm}
  C.~Germani and A.~Kehagias,
  {\it{New Model of Inflation with Non-minimal Derivative Coupling of Standard 
Model Higgs Boson to Gravity}},
  Phys.\ Rev.\ Lett.\  {\bf 105}, 011302 (2010)
  [\href{https://arxiv.org/abs/1003.2635}{{\tt arXiv:1003.2635}}].

\bibitem{Feng:2010ya}
  C.~J.~Feng, X.~Z.~Li and E.~N.~Saridakis,
  {\it{Preventing eternality in phantom inflation}},
  Phys.\ Rev.\ D  {\bf 82}, 023526 (2010)
  [\href{https://arxiv.org/abs/1004.1874}{{\tt arXiv:1004.1874}}].

\bibitem{Hertzberg:2010dc}
  M.~P.~Hertzberg,
  {\it{On Inflation with Non-minimal Coupling}},
  JHEP  {\bf 11}, 023 (2010)
  [\href{https://arxiv.org/abs/1002.2995}{{\tt arXiv:1002.2995}}].

  
  
  
  
\bibitem{Hossain:2014xha}
  M.~W.~Hossain, R.~Myrzakulov, M.~Sami and E.~N.~Saridakis,
  {\it{Variable gravity: A suitable framework for quintessential inflation}},
  Phys.\ Rev.\ D  {\bf 90}, 023512 (2014)
  [\href{https://arxiv.org/abs/1402.6661}{{\tt arXiv:1402.6661}}].

\bibitem{Hossain:2014coa}
  M.~W.~Hossain, R.~Myrzakulov, M.~Sami and E.~N.~Saridakis,
  {\it{Class of quintessential inflation models with parameter space consistent 
with BICEP2}},
  Phys.\ Rev.\ D  {\bf 89}, 123513 (2014)
  [\href{https://arxiv.org/abs/1404.1445}{{\tt arXiv:1404.1445}}].

  
  
  \bibitem{Nojiri:2015fra}
  S.~Nojiri, S.~D.~Odintsov and V.~K.~Oikonomou,
  {\it{Quantitative analysis of singular inflation with scalar-tensor and 
modified gravity}},
  Phys.\ Rev.\ D  {\bf 91}, 084059 (2015)
  [\href{https://arxiv.org/abs/1502.07005}{{\tt arXiv:1502.07005}}].
  
  
  %\cite{Basilakos:2015sza}
\bibitem{Basilakos:2015sza}
S.~Basilakos and J.~D.~Barrow,
 {\it{Hyperbolic Inflation in the Light of Planck 2015 data,}}
Phys. Rev. D \textbf{91}, 103517 (2015)
  [\href{https://arxiv.org/abs/1504.03469}{{\tt arXiv:1504.03469}}].
  
 

  
\bibitem{Hossain:2014ova}
  M.~W.~Hossain, R.~Myrzakulov, M.~Sami and E.~N.~Saridakis,
  {\it{Evading Lyth bound in models of quintessential inflation}},
  Phys.\ Lett.\ B  {\bf 737}, 191-195 (2014)
  [\href{https://arxiv.org/abs/1405.7491}{{\tt arXiv:1405.7491}}].

  
  
  
\bibitem{Geng:2017mic}
  C.~Q.~Geng, C.~C.~Lee, M.~Sami, E.~N.~Saridakis and A.~A.~Starobinsky,
  {\it{Observational constraints on successful model of quintessential 
Inflation}},
  JCAP  {\bf 06}, 011 (2017)
  [\href{https://arxiv.org/abs/1705.01329}{{\tt arXiv:1705.01329}}].
  

\bibitem{Geng:2015fla}
  C.~Q.~Geng, M.~W.~Hossain, R.~Myrzakulov, M.~Sami and E.~N.~Saridakis,
  {\it{Quintessential inflation with canonical and noncanonical scalar fields 
and Planck 2015 results}},
  Phys.\ Rev.\ D  {\bf 92}, 023522 (2015)
  [\href{https://arxiv.org/abs/1502.03597}{{\tt arXiv:1502.03597}}].

  
    \bibitem{Braglia:2020fms}
  M.~Braglia, D.~K.~Hazra, L.~Sriramkumar, and F.~Finelli,
  {\it{Generating primordial features at large scales in two field models of 
inflation}},
  JCAP {\bf 08}, 025 (2020)
  [\href{https://arxiv.org/abs/2004.00672}{{\tt arXiv:2004.00672}}].

  
  
\bibitem{Lola:2020lvk}
  S.~Lola, A.~Lymperis and E.~N.~Saridakis,
  {\it{Inflation with non-canonical scalar fields revisited}},
  Eur.\ Phys.\ J.\ C  {\bf 81}, 719 (2021)
  [\href{https://arxiv.org/abs/2005.14069}{{\tt arXiv:2005.14069}}].
 
\bibitem{Benisty:2019jqz}
  D.~Benisty, E.~I.~Guendelman, E.~N.~Saridakis, H.~Stoecker, J.~Struckmeier 
and 
D.~Vasak,
  {\it{Inflation from fermions with curvature-dependent mass}},
  Phys.\ Rev.\ D  {\bf 100}, 043523 (2019)
  [\href{https://arxiv.org/abs/1905.03731}{{\tt arXiv:1905.03731}}].
  
  

\bibitem{Karydas:2021wmx}
  S.~Karydas, E.~Papantonopoulos and E.~N.~Saridakis,
  {\it{Successful Higgs inflation from combined nonminimal and derivative 
couplings}},
  Phys.\ Rev.\ D  {\bf 104}, 023530 (2021)
  [\href{https://arxiv.org/pdf/2102.08450}{{\tt arXiv:2102.08450}}].

  

  
\bibitem{Panda:2022can}
  S.~Panda, A.~Rana and R.~Thakur,
  {\it{Constant-roll inflation in modified $f(R,\phi )$ gravity model using 
Palatini formalism}},
  Eur.\ Phys.\ J.\ C  {\bf 83}, 297 (2023)
  [\href{https://arxiv.org/abs/2212.00472}{{\tt arXiv:2212.00472}}].
  
\bibitem{Papanikolaou:2022did}
  T.~Papanikolaou, A.~Lymperis, S.~Lola and E.~N.~Saridakis,
  {\it{Primordial black holes and gravitational waves from non-canonical 
inflation}},
  JCAP  {\bf 03}, 003 (2023)
  [\href{https://arxiv.org/abs/2211.14900}{{\tt arXiv:2211.14900}}].

    \bibitem{Sohail:2024oki}
  Sk.~Sohail, S.~Alam, S.~Akthar, and Md.~W.~Hossain,
  {\it{Quintessential early dark energy}},
  Phys.\ Dark.\ Univ.\ {\bf 48}, 101948 (2025)
  [\href{https://arxiv.org/abs/2408.03229}{{\tt arXiv:2408.03229}}].
  
  

\bibitem{Basilakos:2023xof}
  S.~Basilakos, D.~V.~Nanopoulos, T.~Papanikolaou, E.~N.~Saridakis and 
C.~Tzerefos,
  {\it{Gravitational wave signatures of no-scale supergravity in NANOGrav and 
beyond}},
  Phys.\ Lett.\ B\  {\bf 850}, 138507 (2024)
  [\href{https://arxiv.org/abs/2307.08601}{{\tt arXiv:2307.08601}}].
 
  
    
  
  
    
  
\bibitem{Kobayashi:2010cm}
  T.~Kobayashi, M.~Yamaguchi and J.~Yokoyama,
  {\it{G-inflation: Inflation driven by the Galileon field}},
  Phys.\ Rev.\ Lett.\  {\bf 105}, 231302 (2010)
  [\href{https://arxiv.org/abs/1008.0603}{{\tt arXiv:1008.0603}}].

\bibitem{Burrage:2010cu}
  C.~Burrage, C.~de Rham, D.~Seery and A.~J.~Tolley,
  {\it{Galileon inflation}},
  JCAP  {\bf 01}, 014 (2011)
  [\href{https://arxiv.org/abs/1009.2497}{{\tt arXiv:1009.2497}}].

\bibitem{Renaux-Petel:2011lur}
  S.~Renaux-Petel,
  {\it{Orthogonal non-Gaussianities from Dirac-Born-Infeld Galileon inflation}},
  Class.\ Quant.\ Grav.\  {\bf 28}, 182001 (2011)
  [\href{https://arxiv.org/abs/1105.6366}{{\tt arXiv:1105.6366}}].

\bibitem{Renaux-Petel:2011rmu}
  S.~Renaux-Petel, S.~Mizuno and K.~Koyama,
  {\it{Primordial fluctuations and non-Gaussianities from multifield DBI 
Galileon inflation}},
  JCAP  {\bf 11}, 042 (2011)
  [\href{https://arxiv.org/abs/1108.0305}{{\tt arXiv:1108.0305}}].

\bibitem{Choudhury:2012whm}
  S.~Choudhury and S.~Pal,
  {\it{Primordial non-Gaussian features from DBI Galileon inflation}},
  Eur.\ Phys.\ J.\ C  {\bf 75}, 241 (2015)
  [\href{https://arxiv.org/abs/1210.4478}{{\tt arXiv:1210.4478}}].

\bibitem{Ohashi:2012wf}
  J.~Ohashi and S.~Tsujikawa,
  {\it{Potential-driven Galileon inflation}},
  JCAP  {\bf 10}, 035 (2012)
  [\href{https://arxiv.org/abs/1207.4879}{{\tt arXiv:1207.4879}}].


  
\bibitem{Tsujikawa:2014mba}
  S.~Tsujikawa,
  {\it{The effective field theory of inflation/dark energy and the Horndeski 
theory}},
  Lect.\ Notes\ Phys.\  {\bf 892}, 97-136 (2015)
  [\href{https://arxiv.org/abs/1404.2684}{{\tt arXiv:1404.2684}}].

  \bibitem{Qiu:2015nha}
  T.~Qiu and Y.~T.~Wang,
  {\it{G-Bounce Inflation: Towards Nonsingular Inflation Cosmology with 
Galileon 
Field}},
  JHEP  {\bf 04}, 130 (2015)
  [\href{https://arxiv.org/abs/1501.03568}{{\tt arXiv:1501.03568}}].

  
\bibitem{Germani:2015plv}
  C.~Germani, N.~Kudryashova and Y.~Watanabe,
  {\it{On post-inflation validity of perturbation theory in Horndeski 
scalar-tensor models}},
  JCAP  {\bf 08}, 015 (2016)
  [\href{https://arxiv.org/abs/1512.06344}{{\tt arXiv:1512.06344}}].

\bibitem{BeltranJimenez:2017cbn}
  J.~Beltran Jimenez, L.~Heisenberg, R.~Kase, R.~Namba and S.~Tsujikawa,
  {\it{Instabilities in Horndeski Yang-Mills inflation}},
  Phys.\ Rev.\ D  {\bf 95}, 063533 (2017)
  [\href{https://arxiv.org/abs/1702.01193}{{\tt arXiv:1702.01193}}].

\bibitem{Sebastiani:2017cey}
  L.~Sebastiani, S.~Myrzakul and R.~Myrzakulov,
  {\it{Reconstruction of k-essence inflation in Horndeski gravity}},
  Eur.\ Phys.\ J.\ Plus  {\bf 132}, 433 (2017)
  [\href{https://arxiv.org/abs/1702.00064}{{\tt arXiv:1702.00064}}].

\bibitem{Oikonomou:2020sij}
  V.~K.~Oikonomou and F.~P.~Fronimos,
  {\it{Reviving non-minimal Horndeski-like theories after GW170817: kinetic 
coupling corrected Einstein-Gauss-Bonnet inflation}},
  Class.\ Quant.\ Grav.\  {\bf 38}, 035013 (2021)
  [\href{https://arxiv.org/abs/2006.05512}{{\tt arXiv:2006.05512}}].

\bibitem{Chen:2021nio}
  P.~Chen, S.~Koh and G.~Tumurtushaa,
  {\it{Primordial black holes and induced gravitational waves from inflation in 
the Horndeski theory of gravity}},
  [\href{https://arxiv.org/abs/2107.08638}{{\tt arXiv:2107.08638}}].

  
  
  
\bibitem{Choudhury:2023hvf}
  S.~Choudhury, S.~Panda and M.~Sami,
  {\it{Galileon inflation evades the no-go for PBH formation in the 
single-field 
framework}},
  JCAP  {\bf 08}, 078 (2023)
  [\href{https://arxiv.org/abs/2304.04065}{{\tt arXiv:2304.04065}}].

\bibitem{Choudhury:2023hfm}
  S.~Choudhury, A.~Karde, S.~Panda and M.~Sami,
  {\it{Scalar induced gravity waves from ultra slow-roll galileon inflation}},
  Nucl.\ Phys.\ B  {\bf 1007}, 116678 (2024)
  [\href{https://arxiv.org/abs/2308.09273}{{\tt arXiv:2308.09273}}].
 
  
\bibitem{Barrow:1988xh}
  J.~D.~Barrow and S.~Cotsakis,
  {\it{Inflation and the Conformal Structure of Higher Order Gravity Theories}},
  Phys.\ Lett.\ B  {\bf 214}, 515-518 (1988).

   
\bibitem{Garcia-Bellido:1995him}
  J.~Garcia-Bellido and D.~Wands,
  {\it{Constraints from inflation on scalar - tensor gravity theories}},
  Phys.\ Rev.\ D  {\bf 52}, 6739 (1995)
  [\href{https://arxiv.org/abs/gr-qc/9506050}{{\tt arXiv:gr-qc/9506050}}].

  
  

\bibitem{Nojiri:2003ft}
  S.~Nojiri and S.~D.~Odintsov,
  {\it{Modified gravity with negative and positive powers of the curvature: 
Unification of the inflation and of the cosmic acceleration}},
  Phys.\ Rev.\ D  {\bf 68}, 123512 (2003)
  [\href{https://arxiv.org/abs/hep-th/0307288}{{\tt arXiv:hep-th/0307288}}].

 
  
  
  
\bibitem{Carter:2005fu}
  B.~M.~Carter and I.~P.~Neupane,
  {\it{Towards inflation and dark energy cosmologies from modified Gauss-Bonnet 
theory}},
  JCAP  {\bf 06}, 004 (2006)
  [\href{https://arxiv.org/abs/hep-th/0512262}{{\tt arXiv:hep-th/0512262}}].

\bibitem{Ferraro:2006jd}
  R.~Ferraro and F.~Fiorini,
  {\it{Modified teleparallel gravity: Inflation without inflaton}},
  Phys.\ Rev.\ D  {\bf 75}, 084031 (2007)
  [\href{https://arxiv.org/abs/gr-qc/0610067}{{\tt arXiv:gr-qc/0610067}}].

\bibitem{Nojiri:2007as}
  S.~Nojiri and S.~D.~Odintsov,
  {\it{Unifying inflation with LambdaCDM epoch in modified f(R) gravity 
consistent with Solar System tests}},
  Phys.\ Lett.\ B  {\bf 657}, 238-245 (2007)
  [\href{https://arxiv.org/abs/0707.1941}{{\tt arXiv:0707.1941}}].

\bibitem{Nojiri:2007cq}
  S.~Nojiri and S.~D.~Odintsov,
  {\it{Modified f(R) gravity unifying R**m inflation with Lambda CDM epoch}},
  Phys.\ Rev.\ D  {\bf 77}, 026007 (2008)
  [\href{https://arxiv.org/abs/0710.1738}{{\tt arXiv:0710.1738}}].

\bibitem{Mukohyama:2009gg}
  S.~Mukohyama,
  {\it{Scale-invariant cosmological perturbations from Horava-Lifshitz gravity 
without inflation}},
  JCAP  {\bf 06}, 001 (2009)
  [\href{https://arxiv.org/abs/0904.2190}{{\tt arXiv:0904.2190}}].

  
  
  \bibitem{Nojiri:2007bt}
  S.~Nojiri, S.~D.~Odintsov and P.~V.~Tretyakov,
  {\it{From inflation to dark energy in the non-minimal modified gravity}},
  Prog.\ Theor.\ Phys.\ Suppl.\  {\bf 172}, 81-89 (2008)
  [\href{https://arxiv.org/abs/0710.5232}{{\tt arXiv:0710.5232}}].

\bibitem{Bamba:2008ja}
  K.~Bamba and S.~D.~Odintsov,
  {\it{Inflation and late-time cosmic acceleration in non-minimal Maxwell-F(R) 
gravity and the generation of large-scale magnetic fields}},
  JCAP  {\bf 04}, 024 (2008)
  [\href{https://arxiv.org/abs/0801.0954}{{\tt arXiv:0801.0954}}].


  
  
\bibitem{Nozari:2010kri}
  K.~Nozari and B.~Fazlpour,
  {\it{Non-Minimal Warm Inflation and Perturbations on the Warped DGP Brane 
with 
Modified Induced Gravity}},
  Gen.\ Rel.\ Grav.\  {\bf 43}, 207-234 (2011)
  [\href{https://arxiv.org/abs/0906.5047}{{\tt arXiv:0906.5047}}].

\bibitem{Elizalde:2010ep}
  E.~Elizalde, S.~Nojiri, S.~D.~Odintsov and D.~Saez-Gomez,
  {\it{Unifying inflation with dark energy in modified F(R) Horava-Lifshitz 
gravity}},
  Eur.\ Phys.\ J.\ C  {\bf 70}, 351-361 (2010)
  [\href{https://arxiv.org/abs/1006.3387}{{\tt arXiv:1006.3387}}].

\bibitem{Cai:2010kp}
  Y.~F.~Cai and E.~N.~Saridakis,
  {\it{Inflation in Entropic Cosmology: Primordial Perturbations and 
non-Gaussianities}},
  Phys.\ Lett.\ B  {\bf 697}, 280-287 (2011)
  [\href{https://arxiv.org/abs/1011.1245}{{\tt arXiv:1011.1245}}].

\bibitem{Briscese:2012ys}
  F.~Briscese, A.~Marcian{\`o}, L.~Modesto and E.~N.~Saridakis,
  {\it{Inflation in (Super-)renormalizable Gravity}},
  Phys.\ Rev.\ D  {\bf 87}, 083507 (2013)
  [\href{https://arxiv.org/abs/1212.3611}{{\tt arXiv:1212.3611}}].

\bibitem{Sebastiani:2013eqa}
  L.~Sebastiani, G.~Cognola, R.~Myrzakulov, S.~D.~Odintsov and S.~Zerbini,
  {\it{Nearly Starobinsky inflation from modified gravity}},
  Phys.\ Rev.\ D  {\bf 89}, 023518 (2014)
  [\href{https://arxiv.org/abs/1311.0744}{{\tt arXiv:1311.0744}}].

\bibitem{Bamba:2014jia}
  K.~Bamba, R.~Myrzakulov, S.~D.~Odintsov and L.~Sebastiani,
  {\it{Trace-anomaly driven inflation in modified gravity and the BICEP2 
result}},
  Phys.\ Rev.\ D  {\bf 90}, 043505 (2014)
  [\href{https://arxiv.org/abs/1403.6649}{{\tt arXiv:1403.6649}}].

\bibitem{Bamba:2014mua}
  K.~Bamba, G.~Cognola, S.~D.~Odintsov and S.~Zerbini,
  {\it{One-loop modified gravity in a de Sitter universe, quantum-corrected 
inflation, and its confrontation with the Planck result}},
  Phys.\ Rev.\ D  {\bf 90}, 023525 (2014)
  [\href{https://arxiv.org/abs/1404.4311}{{\tt arXiv:1404.4311}}].

\bibitem{Nojiri:2014zqa}
  S.~Nojiri and S.~D.~Odintsov,
  {\it{Mimetic $F(R)$ gravity: inflation, dark energy and bounce}},
  [\href{https://arxiv.org/abs/1408.3561}{{\tt arXiv:1408.3561}}].

  

  

\bibitem{DeLaurentis:2015fea}
  M.~De Laurentis, M.~Paolella and S.~Capozziello,
  {\it{Cosmological inflation in $F(R,\mathcal{G})$ gravity}},
  Phys.\ Rev.\ D  {\bf 91}, 083531 (2015)
  [\href{https://arxiv.org/abs/1503.04659}{{\tt arXiv:1503.04659}}].

\bibitem{Bamba:2016wjm}
  K.~Bamba, S.~D.~Odintsov and E.~N.~Saridakis,
  {\it{Inflationary cosmology in unimodular $F(T)$ gravity}},
  Mod.\ Phys.\ Lett.\ A  {\bf 32}, 1750114 (2017)
  [\href{https://arxiv.org/abs/1605.02461}{{\tt arXiv:1605.02461}}].
 
   

 
\bibitem{Paliathanasis:2017apr}
A.~Paliathanasis,
 {\it{Analytic Solution of the Starobinsky Model for Inflation}},
Eur. Phys. J. C \textbf{77}, no.7, 438 (2017)
  [\href{https://arxiv.org/abs/1706.06400}{{\tt arXiv:1706.06400}}].
 



\bibitem{Nojiri:2017ncd}
  S.~Nojiri, S.~D.~Odintsov and V.~K.~Oikonomou,
  {\it{Modified Gravity Theories on a Nutshell: Inflation, Bounce and Late-time 
Evolution}},
  Phys.\ Rept.\  {\bf 692}, 1-104 (2017)
  [\href{https://arxiv.org/abs/1705.11098}{{\tt arXiv:1705.11098}}].

\bibitem{Awad:2017ign}
  A.~Awad, W.~El Hanafy, G.~G.~L.~Nashed, S.~D.~Odintsov and V.~K.~Oikonomou,
  {\it{Constant-roll Inflation in $f(T)$ Teleparallel Gravity}},
  JCAP  {\bf 07}, 026 (2018)
  [\href{https://arxiv.org/abs/1710.00682}{{\tt arXiv:1710.00682}}].

\bibitem{Chakraborty:2018scm}
  S.~Chakraborty, T.~Paul and S.~SenGupta,
  {\it{Inflation driven by Einstein-Gauss-Bonnet gravity}},
  Phys.\ Rev.\ D  {\bf 98}, 083539 (2018)
  [\href{https://arxiv.org/abs/1804.03004}{{\tt arXiv:1804.03004}}].

\bibitem{Qiu:2018nle}
  T.~Qiu, K.~Tian and S.~Bu,
  {\it{Perturbations of bounce inflation scenario from $f(T)$ modified gravity 
revisited}},
  Eur.\ Phys.\ J.\ C  {\bf 79}, 261 (2019)
  [\href{https://arxiv.org/abs/1810.04436}{{\tt arXiv:1810.04436}}].

\bibitem{Shimada:2018lnm}
  K.~Shimada, K.~Aoki and K.~i.~Maeda,
  {\it{Metric-affine Gravity and Inflation}},
  Phys.\ Rev.\ D  {\bf 99}, 104020 (2019)
  [\href{https://arxiv.org/abs/1812.03420}{{\tt arXiv:1812.03420}}].
  
  


\bibitem{Gamonal:2020itt}
  M.~Gamonal,
  {\it{Slow-roll inflation in $f(R,T)$ gravity and a modified Starobinsky-like 
inflationary model}},
  Phys.\ Dark\ Univ.\  {\bf 31}, 100768 (2021)
  [\href{https://arxiv.org/abs/2010.03861}{{\tt arXiv:2010.03861}}].

    
\bibitem{CANTATA:2021ktz}
  E.~N.~Saridakis {\it{et al.}} [CANTATA],
  {\it{Modified Gravity and Cosmology: An Update by the CANTATA Network}},
  [\href{https://arxiv.org/abs/2105.12582}{{\tt arXiv:2105.12582}}].

  
\bibitem{Cai:2021uup}
  R.~G.~Cai, C.~Fu and W.~W.~Yu,
  {\it{Parity violation in stochastic gravitational wave background from 
inflation in Nieh-Yan modified teleparallel gravity}},
  Phys.\ Rev.\ D  {\bf 105}, 103520 (2022)
  [\href{https://arxiv.org/abs/2112.04794}{{\tt arXiv:2112.04794}}].

\bibitem{Papanikolaou:2021uhe}
  T.~Papanikolaou, C.~Tzerefos, S.~Basilakos and E.~N.~Saridakis,
  {\it{Scalar induced gravitational waves from primordial black hole Poisson 
fluctuations in f(R) gravity}},
  JCAP  {\bf 10}, 013 (2022)
  [\href{https://arxiv.org/abs/2112.15059}{{\tt arXiv:2112.15059}}].

\bibitem{Shiravand:2022ccb}
  M.~Shiravand, S.~Fakhry and M.~Farhoudi,
  {\it{Cosmological inflation in f(Q,T) gravity}},
  Phys.\ Dark\ Univ.\  {\bf 37}, 101106 (2022)
  [\href{https://arxiv.org/abs/2204.00906}{{\tt arXiv:2204.00906}}].


\bibitem{Capozziello:2022tvv}
  S.~Capozziello and M.~Shokri,
  {\it{Slow-roll inflation in f(Q) non-metric gravity}},
  Phys.\ Dark\ Univ.\  {\bf 37}, 101113 (2022)
  [\href{https://arxiv.org/abs/2209.06670}{{\tt arXiv:2209.06670}}].

   
   
\bibitem{Tzerefos:2023mpe}
  C.~Tzerefos, T.~Papanikolaou, E.~N.~Saridakis and S.~Basilakos,
  {\it{Scalar induced gravitational waves in modified teleparallel gravity 
theories}},
  Phys.\ Rev.\ D\  {\bf 107}, 124019 (2023)
  [\href{https://arxiv.org/abs/2303.16695}{{\tt arXiv:2303.16695}}].

 

\bibitem{Bhat:2023qwa}
  A.~Bhat, S.~Mandal and P.~K.~Sahoo,
  {\it{Slow-roll inflation in$f(T,\mathcal{T})$ modified gravity}},
  Chin.\ Phys.\ C  {\bf 47}, 125104 (2023)
  [\href{https://arxiv.org/abs/2310.05987}{{\tt arXiv:2310.05987}}].

\bibitem{Sadatian:2024lub}
  S.~D.~Sadatian and S.~M.~R.~Hosseini,
  {\it{Cosmological inflation in the modified gravity model f(Q,C)}},
  Phys.\ Dark\ Univ.\  {\bf 47}, 101737 (2025)


  
  
    
    \bibitem{Starobinsky:1980te}
  A.~A.~Starobinsky,
  {\it{A New Type of Isotropic Cosmological Models Without Singularity}},
  Phys.\ Lett.\ B {\bf 91}, 99 (1980).

  
  
  
  
  
  
  
  
  
  
  
  
  
  
  
  
  

 

\bibitem{Kallosh:2013hoa}
  R.~Kallosh and A.~Linde,
  {\it{Universality Class in Conformal Inflation}},
  JCAP  {\bf 07}, 002 (2013)
  [\href{https://arxiv.org/abs/1306.5220}{{\tt arXiv:1306.5220}}].

\bibitem{Kallosh:2013yoa}
  R.~Kallosh, A.~Linde and D.~Roest,
  {\it{Superconformal Inflationary $\alpha$-Attractors}},
  JHEP  {\bf 11}, 198 (2013)
  [\href{https://arxiv.org/abs/1311.0472}{{\tt arXiv:1311.0472}}].

\bibitem{Kallosh:2013tua}
  R.~Kallosh, A.~Linde and D.~Roest,
  {\it{Universal Attractor for Inflation at Strong Coupling}},
  Phys.\ Rev.\ Lett.\  {\bf 112}, 011303 (2014)
  [\href{https://arxiv.org/abs/1310.3950}{{\tt arXiv:1310.3950}}].

\bibitem{Galante:2014ifa}
  M.~Galante, R.~Kallosh, A.~Linde and D.~Roest,
  {\it{Unity of Cosmological Inflation Attractors}},
  Phys.\ Rev.\ Lett.\  {\bf 114}, 141302 (2015)
  [\href{https://arxiv.org/abs/1412.3797}{{\tt arXiv:1412.3797}}].

\bibitem{Linde:2015uga}
  A.~Linde,
  {\it{Single-field $\alpha$-attractors}},
  JCAP  {\bf 05}, 003 (2015)
  [\href{https://arxiv.org/abs/1504.00663}{{\tt arXiv:1504.00663}}].

  
  
   
  
  
  
\bibitem{Gao:2017uja}
  Q.~Gao and Y.~Gong,
  {\it{Reconstruction of extended inflationary potentials for attractors}},
  Eur.\ Phys.\ J.\ Plus  {\bf 133}, 491 (2018)
  [\href{https://arxiv.org/abs/1703.02220}{{\tt arXiv:1703.02220}}].

\bibitem{Miranda:2017juz}
  T.~Miranda, J.~C.~Fabris and O.~F.~Piattella,
  {\it{Reconstructing a $f(R)$ theory from the $\alpha$-Attractors}},
  JCAP  {\bf 09}, 041 (2017)
  [\href{https://arxiv.org/abs/1707.06457}{{\tt arXiv:1707.06457}}].

\bibitem{Dimopoulos:2017zvq}
  K.~Dimopoulos and C.~Owen,
  {\it{Quintessential Inflation with $\alpha$-attractors}},
  JCAP  {\bf 06}, 027 (2017)
  [\href{https://arxiv.org/abs/1703.00305}{{\tt arXiv:1703.00305}}].

\bibitem{Dimopoulos:2017tud}
  K.~Dimopoulos, L.~Donaldson Wood and C.~Owen,
  {\it{Instant preheating in quintessential inflation with 
$\alpha$-attractors}},
  Phys.\ Rev.\ D  {\bf 97}, 063525 (2018)
  [\href{https://arxiv.org/abs/1712.01760}{{\tt arXiv:1712.01760}}].

\bibitem{Scalisi:2018eaz}
  M.~Scalisi and I.~Valenzuela,
  {\it{Swampland distance conjecture, inflation and $\alpha$-attractors}},
  JHEP  {\bf 08}, 160 (2019)
  [\href{https://arxiv.org/abs/1812.07558}{{\tt arXiv:1812.07558}}].

\bibitem{Braglia:2020bym}
  M.~Braglia, W.~T.~Emond, F.~Finelli, A.~E.~Gumrukcuoglu and K.~Koyama,
  {\it{Unified framework for early dark energy from $\alpha$-attractors}},
  Phys.\ Rev.\ D  {\bf 102}, 083513 (2020)
  [\href{https://arxiv.org/abs/2005.14053}{{\tt arXiv:2005.14053}}].

\bibitem{Bhattacharya:2022akq}
  S.~Bhattacharya, K.~Dutta, M.~R.~Gangopadhyay and A.~Maharana,
  {\it{$\alpha$-attractor inflation: Models and predictions}},
  Phys.\ Rev.\ D  {\bf 107}, 103530 (2023)
  [\href{https://arxiv.org/abs/2212.13363}{{\tt arXiv:2212.13363}}].

\bibitem{Brissenden:2023yko}
  L.~Brissenden, K.~Dimopoulos and S.~S{\'a}nchez L{\'o}pez,
  {\it{Non-oscillating early dark energy and quintessence from 
$\alpha$-attractors}},
  Astropart.\ Phys.\  {\bf 157}, 102925 (2024)
  [\href{https://arxiv.org/abs/2301.03572}{{\tt arXiv:2301.03572}}].

\bibitem{Ferrara:2016fwe}
  S.~Ferrara and R.~Kallosh,
  {\it{Seven-disk manifold, $\alpha$-attractors, and $B$ modes}},
  Phys.\ Rev.\ D  {\bf 94}, 126015 (2016)
  [\href{https://arxiv.org/abs/1610.04163}{{\tt arXiv:1610.04163}}].

\bibitem{Akarsu:2016qhf}
  {\"O}.~Akarsu, S.~Boran, E.~O.~Kahya, N.~{\"O}zdemir and M.~Ozkan,
  {\it{Broken scale invariance, $\alpha$-attractors and vector impurity}},
  Eur.\ Phys.\ J.\ C  {\bf 77}, 306 (2017)
  [\href{https://arxiv.org/abs/1606.05308}{{\tt arXiv:1606.05308}}].
  
%\cite{ACT:2025tim}
\bibitem{ACT:2025tim}
E.~Calabrese \textit{et al.} [ACT],
  {\it{The Atacama Cosmology Telescope: DR6 Constraints on Extended 
Cosmological Models,}}
  [\href{https://arxiv.org/abs/2503.14454}{{\tt arXiv:2503.14454}}].
  
    

 

 
\end{thebibliography}
\end{document}